\documentclass[9pt,twocolumn,twoside]{opticajnl}
\journal{opticajournal} 

\setboolean{shortarticle}{true}

\usepackage{lineno}

\usepackage{siunitx}

\newcommand{\rs}{\scriptscriptstyle} 
\newcommand{\fS}{f_{\rs S}} 
\newcommand{\fAOM}{f_{\rs AOM}} 
\newcommand{\fL}{f_{\rs L}} 
\newcommand{\phiXT}{\phi_{\rs \mathrm{XT}}} 
\newcommand{\phiTh}{\phi_{\rs \theta}} 
\newcommand{\deltafAOMvalueI}{\SI{9.41}{\mega\hertz}}

\title{High-Precision Frequency-Controlled Optical Phase Shifter with Acousto Optic Devices}

\author[1]{Eduardo Esquivel-Ramírez}
\author[1]{Leonardo Uhthoff-Rodríguez}
\author[1]{Edgar Giovanni Alonso-Torres}
\author[1]{Alberto Hernández-López}
\author[1]{Carlos Gardea-Flores}
\author[1,*]{Asaf Paris-Mandoki}

\affil[1]{Instituto de Física, Universidad Nacional Autónoma de México, Cd. de México C.P. 04510, México}

\affil[*]{asaf@fisica.unam.mx}

\begin{abstract}
A fundamental parameter to determine how electromagnetic waves interfere is their relative phase, and achieving a fine control over it enables a wide range of interferometric applications. Existing phase control methods rely on modifying the optical path length by either changing the path followed by the light or by altering the thickness or index of refraction of an optical element in the setup. In this work we present a novel method, based on acousto-optic modulators (AOM), which allows adjusting the phase by shifting the frequency of the light in a segment of its path. Since the amount of phase-shift depends on the length of the segment, an optical fiber is used to realize a \boldmath$2\pi$ shift. Two experimental implementations are described which deal with different sources of phase fluctuations.  The first addresses fluctuations resulting from the optical fiber while the second also tackles unwanted variations originating from the AOMs.
\end{abstract}

\setboolean{displaycopyright}{false} 

\renewcommand{\section}[1]{} 

\begin{document}

\maketitle

Controlling the phase of light is challenging since it usually requires the ability to adjust the optical path length by a small fraction of an optical wavelength, typically in the order of a few nanometers. In spite of  these difficulties, achieving control of the phase is of great importance as it finds applications in a large number of fields including holographic~\cite{powellInterferometricVibrationAnalysis1965} and phase-shifting~\cite{bruningDigitalWavefrontMeasuring1974} inteferometry as well as homodyne detection in quantum optics~\cite{collettQuantumTheoryOptical1987, banaszekDirectMeasurementWigner1999} and quantum information processing~\cite{kokLinearOpticalQuantum2007}  among others. 

Several methods can be used for controlling phase shifts, including opto-mechanical, thermo-optical~\cite{liuThermoopticPhaseShifters2022} or electro-optical~\cite{sinatkasElectroopticModulationIntegrated2021} which can be easy to implement and have benefited from decades of engineering. Another option is to use acousto-optical means~\cite{liOpticalPhaseShifting2005, ehrlichVoltagecontrolledAcoustoopticPhase1988} where a beam passes through a pair of acousto-optic modulators and the phase difference between the radio frequency (RF) signals driving them is used to control the phase of the light. This method effectively displaces the traveling diffraction grating created by the sound waves in the crystal and thus the optical path length is changed.

In this work we propose and demonstrate a novel acousto-optic method for optical phase shifting controlled by the acoustic frequency. As the frequency of the electrical signal, which generates the acoustic wave, can be controlled with a precision of \SI{1}{\micro\hertz} using a common \SI{100}{\mega\hertz} signal generator, this endows the method of phase tuning with a high level of precision. As opposed to other methods, the method presented in this letter does not rely on modifying the optical path length.

\begin{figure}[ht!]
\centering
\includegraphics[width=\linewidth]{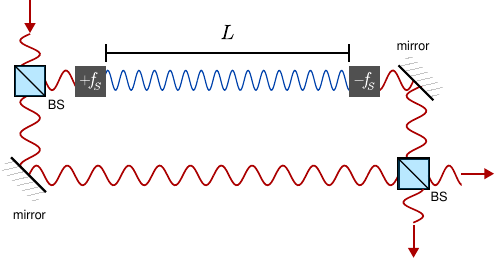}
\caption{Working principle of the phase shifter. A laser beam is divided by a 50/50 beam splitter (BS). One of the branches of the interferometer propagates a distance $L$ with a tunable frequency shift $\fS$. Varying $\fS$ changes the number of wavelengths that fit within $L$. This results in an effective phase shift relative to the non-frequency-shifted branch. Finally both branches are mixed in a BS to analyze the phase shift from the change in the interference pattern. }
\label{fig:mach-zehnder}
\end{figure}

The basic principle of operation of our proposed phase-control method is illustrated in Fig.~\ref{fig:mach-zehnder}. In the Mach-Zehnder interferometer shown, the phase of one arm is shifted with respect to the other. This is accomplished by, first, shifting its frequency by $\fS$, then allowing the light to propagate through a distance $L$, and finally shifting the frequency back by $\fS$ before combining both arms in a beam splitter. Since the arm length is fixed the number of wavelengths that fit within $L$ changes when varying $\fS$. In fact, the additional phase $\phi$  acquired due to the frequency shift is $\phi = 2\pi n L \fS / c$, where $c$ is the speed of light and $n$ the index of refraction of the medium. Therefore, the phase shift range $\Delta\phi$ obtained by changing the modulation frequency by an amount $\Delta\fS$ is
\begin{equation}
    \Delta\phi = \frac{2\pi}{c} n L \Delta\fS .
    \label{eq:PhaseShift}
\end{equation}
By changing the frequency during a fixed length of propagation we achieve a phase shift proportional to the frequency shift instead of the usual case in which the phase shift is proportional to the integral over time of the frequency shift.

To implement the frequency shift we use an acousto-optic modulator (AOM) in a double-pass configuration~\cite{Donley2005} for which $\fS$ can be tuned by a few tens of \si{\mega\hertz} without misalignment. This means that to adjust the  phase shift in a range of $2\pi$ the frequency-shifted path length $L$ must be of several meters. However, a compact setup can be obtained by coupling the frequency shifted arm of the interferometer to an optical fiber. Nevertheless, the phase is highly sensitive to variations in the fiber length, which may result from mechanical and thermal changes. The latter also affects the phase due to alterations in the refractive index of the fiber with temperature \cite{priestThermalCoefficientsRefractive1997}. The resulting phase fluctuation is 
\begin{equation}
\delta\phi = \frac{2\pi}{c}(\fL+\fS)(n\delta L+L\delta n),
\label{eq:PhaseFluctuations1}
\end{equation}
where $\fL$ is the frequency of the light, $\delta L$ represents the length fluctuation, and $\delta n$ is the refractive index variation. 

There is a trade-off involved when choosing $L$. On the one hand, according to Eq.~\ref{eq:PhaseShift} the fiber length can be increased to achieve a larger phase shift range $\Delta\phi$ for a given frequency shift $\Delta \fS$, but on the other hand the phase fluctuations $\delta\phi$ increase. In the case of thermal variations, for example, the variation in length of an optical fiber $\delta L$ changes proportionally to its total length $L$. Hence, both terms in Eq.~\ref{eq:PhaseFluctuations1} increase in proportion to $L$. Therefore, the choice of fiber length should be based on the specific requirements of the application. 

Using acousto-optic modulators for frequency shifting may introduce an additional effect where the angle of the diffracted light, and consequently the optical path, depends on the frequency. The resulting phase shift, denoted as $\Delta\phiTh$, due to a change in angle from a frequency change $\Delta\fS$ is highly sensitive to the specific position of the optical components so it is challenging to calculate this dependence a priori. This effect can be suppressed with careful alignment and for a \SI{1}{\milli\meter} misalignment can contribute up to $10\%$ of the total phase shift. However, despite this sensitivity, the contribution is repeatable because of the one-to-one correspondence between $\phiTh$ and $\fS$. Hence, calibrating the phase shifter adequately addresses this effect.

Another factor contributing to phase variation is the phase fluctuation $\delta\phiXT$, arising from changes in the width of the AOMs' crystals due to thermal effects. The thermal origin of this variation poses challenges in achieving precise control over it. Notably, conventional AOM drivers do not exhibit a perfectly flat power output as a function of  frequency, leading to fluctuations in the power delivered to the crystal. These fluctuations, in turn, cause unintended changes in the crystal's width which may lead to systematic errors in the phase shift. 

\begin{figure}[t]
\centering
\includegraphics[width=\linewidth]{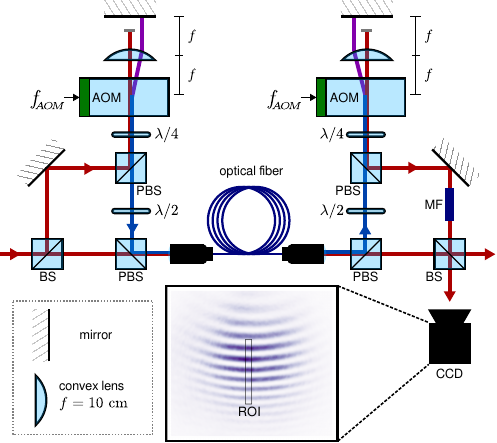}
\caption{Experimental setup which suppresses fiber related phase fluctuations. A \SI{780}{\nano\meter} linearly polarized laser beam is divided by a 50/50 beam splitter (BS). One arm goes through an AOM in a double-pass configuration to change its frequency before going through a \SI{10}{\meter} single-mode optical fiber with an opposite polarization with respect to the other arm. The $\lambda /4$ waveplate enables the transmission of the returning beam through the polarizing beam splitter (PBS), while a $\lambda /2$ waveplate allows for further reflection of this beam at the subsequent PBS. After going through the fiber a PBS separates the beams and the frequency shifted arm goes through a second double-pass AOM to restore its original frequency. A single-mode optical fiber acts as a mode filter (MF) of the shifted branch, then both arms interfere in a BS. A CCD camera captures the interference pattern and we select a region of interest (ROI) for analysis.}
\label{fig:setup}
\end{figure}

\begin{figure}[ht]
\centering
\includegraphics[width=\linewidth]{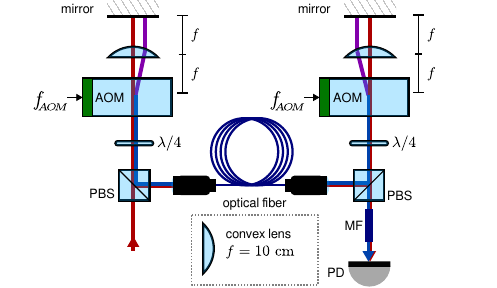}
\caption{Experimental setup for suppressing the phase fluctuations caused by the fiber and by thermal expansion of the AOMs' crystals. The zero-order beam of the AOM acts as the non-shifted branch of the interferometer. Both arms travel through a \SI{10}{\meter} single-mode optical fiber before passing a second AOM to restore the frequency of the shifted beam. A single-mode optical fiber acts as a mode filter (MF) and a photodiode (PD) measures the intensity at the output of the interferometer to analyze the phase. }
\label{fig:setupV2}
\end{figure}

To demonstrate the working principle of the phase shifter we assemble two different Mach-Zehnder interferometers. The first one shown in Fig.~\ref{fig:setup} addresses the phase fluctuation due to fiber length variations. The second one (Fig.~\ref{fig:setupV2}) also suppresses the AOM crystal thermal expansion effects. 


The unwanted fiber fluctuations can be reduced by noting that the main contribution presented in Eq.~\ref{eq:PhaseFluctuations1} comes from the term with $\fL$, which is more than six orders of magnitude larger than $\fS$. Therefore, a straightforward solution to mitigate their effects is to couple both arms of the interferometer into the same fiber and separate them at the output. This results in a strong suppression since, instead of behaving according to Eq.~\ref{eq:PhaseFluctuations1}, the terms with $\fL$ cancel out and the phase fluctuation due to variations in fiber length or refractive index is given by 
\begin{equation}
\delta\phi = \frac{2\pi}{c}\fS(n\delta L+L\delta n).
\label{eq:PhiFluctuations}
\end{equation}
Intuitively, if the two arms of the interferometer traverse identical paths, they are subject to the same fluctuations, thus maintaining a constant relative phase difference.

This reasoning motivates the interferometer proposed in Fig.~\ref{fig:setup}, both arms are folded on top of each other with opposite polarization and coupled into a \SI{10}{\meter} long polarization-maintaining optical fiber. The frequency of one arm is shifted by $\fS=2\fAOM$ using an AOM in a double-pass configuration \cite{Donley2005}, with $\fAOM$ the AOM frequency, allowing us to scan $\fS$ while maintaining the fiber coupling. A second AOM in a double-pass configuration subtracts the frequency shift of $\fS$. Finally, we clean the mode of the frequency-shifted beam with a mode filter (MF) consisting of a single-mode optical fiber before making it interfere.

We can estimate the frequency change required to produce phase shift of $\Delta\phi=2\pi$ in this setup. An important difference between the ideal setup of Fig.~\ref{fig:mach-zehnder} and the actual experimental setup from Fig.~\ref{fig:setup} is that as we are using the AOMs in a double-pass configuration the frequency shifted beam travels some distance with an added frequency $\fAOM$ and other distance with $2\fAOM$. Measuring all the optical paths to determine $L$ we estimate that $\Delta\fAOM \approx \SI{9.24}{\mega\hertz}$ for a $2\pi$ phase shift. However, this calculation only accounts for the phase shift due to the distance propagated with different frequency and to have a quantitative agreement with the measured phase shift also $\delta\phiXT$ and $\Delta\phiTh$ should be accounted for.


\begin{figure}[ht]
\centering
\includegraphics[width=\linewidth]{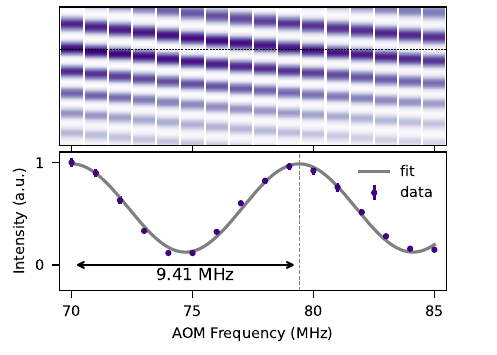}
\caption{(top) Interferometric fringes displaced due to the phase shift introduced by the experimental setup shown in Fig.~\ref{fig:setup}. As the AOM frequency is scanned the relative phase between the interferometer branches changes moving the interference pattern. (bottom) Intensity over a horizontal cut shown with dashed line in the top image. The gray line is a sinusoidal fit and the vertical dashed line shows the $2\pi$ phase shift with $\Delta\fAOM = \SI{9.41}{\mega\hertz}$ obtained form the fit.}
\label{fig:results}
\end{figure}

To observe the phase shift we systematically varied the frequency $\fAOM$ by changing the output frequency of two direct digital synthesizers (DDS) channels that feed the AOMs while capturing images of the interference pattern. 
We select a region of interest (ROI) from the captured images to facilitate comparison of the phase in such a way that the interference pattern is clear and show numerous fringes, the size of the ROI will depend in the size of the light modes at the camera position. We display the interference fringes as a function of $\fAOM$ in Fig.~\ref{fig:results}. During this process, we incrementally adjusted the frequency from \SI{70}{\mega\hertz} to \SI{85}{\mega\hertz} with a $\SI{15}{\milli\second}$ interval between each measurement, which was the smallest delay our setup would allow, in order to avoid the effect of slow drifts of the phase.


Scanning the acoustic frequency results in a displacement of the interference pattern as the phase of one of the arms of the interferometer changes with respect to the other. By tracking the intensity at a fixed position in the image as a function of $\fAOM$ we get that a $2\pi$ phase shift requires a frequency shift $\Delta\fAOM=\deltafAOMvalueI$. However, it is important to note that this required frequency shift may exhibit variations of up to $\sim\SI{0.5}{\mega\hertz}$ depending on whether it is a one-off measurement or part of a continuous measurement for which a stationary state has been reached.  We attribute this effect to the frequency-dependent power delivered by AOM drivers and amplifiers which may cause transient heating or cooling of the AOMs' crytals changing the effective optical path introducing a stray phase shift. In particular the high refractive index of the AOM crystal ($\mathrm{TeO}_{2}$) introduces high variations as it undergoes thermal expansion.

The presented phase shifter can be improved to increase its precision and repeatability by reducing the effect of the AOM crystal thermal fluctuations. We use the same principle of combining both arms of the interferometer into the same path. In this improved case both arms of the interferometer are sent through the AOMs with zero-order acting as the non-shifted branch of the interferometer while the diffracted beam is used as the shifted branch as shown in Fig.~\ref{fig:setupV2}. This scenario suppresses the relative optical path fluctuations from $\delta L$ and $\delta n$ in Eq. \ref{eq:PhiFluctuations} and by an analogous argument $\delta\phiXT$ is also significantly decreased. However, a drawback of this method is that there are additional frequencies $\fL+\fS$ and $\fL-\fS$ together with the outputs at the original frequency $\fL$. The net effect of the extra frequencies is the presence of a beat, at frequencies $\fS$ and $2\fS$, which is averaged with a low-pass filter and appears in the signal as a constant intensity offset. We can analyze the effect of this via the visibility \cite{Saleh}:
\begin{equation}
    \mathcal{V}=\frac{I_{\rs \mathrm{max}}-I_{\rs \mathrm{min}}}{I_{\rs \mathrm{max}}+I_{\rs \mathrm{min}}} .
\label{eq:usualvisibility}
\end{equation}
Where $I_{\rs \mathrm{max}}$ and $I_{\rs \mathrm{min}}$ represent the intensity at constructive and destructive interference respectively. To understand what are the limits on the visibility we can achieve in our setup we calculate the total intensity that will be measured taking into account the interference as
\begin{equation}
I=I_{\rs 0,0}+I_{\rs 0,-1}+I_{\rs +1,0}+I_{\rs +1,-1}+2\sqrt{I_{\rs 0,0}I_{\rs +1,-1}}\gamma\cos\Delta\phi,
\label{eq:Intensity}
\end{equation}
where $I_{\rs i,j}$ is the intensity of the $i$th order of the first AOM and the $j$th of the second one, and $\gamma$ is the normalized cross-correlation which has a value of 1 for completely correlated waves~\cite{Saleh}. Substituting the maximum and minimum intensity values from Eq. \ref{eq:Intensity} into Eq. \ref{eq:usualvisibility} we find
\begin{equation}
\mathcal{V}=\frac{2\sqrt{I_{\rs 0,0}I_{\rs +1,-1}}\gamma}{I_{\rs 0,0}+I_{\rs 0,-1}+I_{\rs +1,0}+I_{\rs +1,-1}}.
\label{eq:Visibility}
\end{equation}
Thus, the additional frequencies present in the setup of Fig.~\ref{fig:setupV2} appear as $I_{\rs 0,-1}$ and $I_{\rs +1,0}$ leading to a decrease of the maximum achievable visibility. To ensure good visibility in the interference measurement, we fixed the power of the RF signal such that $I_{\rs 0,0} = I_{\rs +1,-1}$ for $\fAOM = \SI{80}{\mega\hertz}$. This does not maintain a balanced power distribution throughout the measured frequency range but its impact can be accounted for with Eq.~\ref{eq:Visibility}.



\begin{figure}[ht]
\centering
\includegraphics[width=\linewidth]{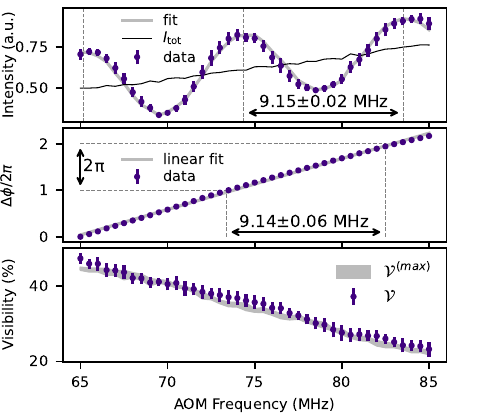}
\caption{(top) Normalized intensity of the interference signal of the setup in Fig.~\ref{fig:setupV2}. The output intensity oscillates as a function of the AOM frequency due to the relative phase being changed. A phase shift $\Delta\phi=2\pi$ is obtained for a frequency change of $\Delta\fAOM=9.15\pm$\SI{ 0.02}{\mega\hertz}. A total of 41 different frequencies, from  \SI{65}{\mega\hertz} to \SI{85}{\mega\hertz}, where measured. The measurements where repeated 23 times and the obtained mean values are plotted. Error bars show the statistical standard deviation. The black line is $I_{\rs\mathrm{tot}}=I_{\rs 0,0}+I_{\rs 0,-1}+I_{\rs +1,0}+I_{\rs +1,-1}$, the sum of the individually measured intensities without interference. The gray line shows a fit acording to Eq.~\ref{eq:Intensity}. (middle) Resulting phase shift as a function of $\fAOM$. The gray line is a linear fit. (bottom) The maximum achievable visibility $\mathcal{V}^{(max)}$ is compared with the actual visibility $\mathcal{V}$ which is calculated using Eq.~\ref{eq:usualvisibility} from the observed $I_{\rs \mathrm{max}}$ and $I_{\rs \mathrm{min}}$ at each AOM frequency.}
\label{fig:resultsV2}
\end{figure}

We obtain an interference signal by recording the intensity using  a DET36A2 photodiode (PD) from Thorlabs while varying $\fAOM$ from \SI{65}{\mega\hertz} to \SI{85}{\mega\hertz}. Since the interfering beams co-propagate in the same spatial mode there is not a changing fringe pattern as in the previous case and the entire pattern varies from bright to dark and the resulting intensity is shown in Fig.~\ref{fig:resultsV2} (top). To measure how the phase shift resulting from our proposed method depends on the AOM frequency, we fix $\fAOM$ and vary the phase of the beams by a separate method, namely by changing the relative phase between the RF signals generated with the DDS~\cite{liOpticalPhaseShifting2005, ehrlichVoltagecontrolledAcoustoopticPhase1988} from $0$ to $2\pi$. This results in a sinusoidal signal for which the initial phase depends on $\fAOM$ and can be determined by fitting the obtained data. This procedure is repeated to obtain the phase for the whole range of frequencies used and the results are shown in Fig. \ref{fig:resultsV2} (middle). A linear relationship is obtained and the deviation from linearity can be attributed to the nonlinear character of $\phiTh$ as a function of $\fAOM$.

To quantify if the proposed method introduces decoherence between the arms, the intensities $I_{\rs 0,0}$, $I_{\rs +1,-1}$, $I_{\rs 0,-1}$ and $I_{\rs +1,0}$ are measured individually to estimate, using Eq. \ref{eq:Visibility}, a theoretical $\mathcal{V}^{(max)}$ for $\gamma=1$ shown in  Fig. \ref{fig:resultsV2} (bottom) as a thick grey line. In the same figure we plot the visibility calculated via Eq. \ref{eq:usualvisibility} where the maximum and minimum values of intensity are found for each value $\fAOM$ by controlling the phase using the relative phase of the RF signals driving the AOMs. The agreement between both sets of data shows that the visibility variation can be solely attributed to intensity changes and is not a consequence of decoherence. 

This optical setup is more robust since most of the relative fluctuations between the interferometer branches are suppressed. For this arrangement we find that $\Delta\fAOM=9.14\pm$\SI{ 0.06}{\mega\hertz} is the calibration frequency which corresponds to a $2\pi$ phase shift. 


We introduced and demonstrated a novel method for implementing a high-precision frequency-controlled phase shifter based on acousto-optic modulators. Two experimental setups were developed to address the different experimental fluctuations that affect the performance of the phase shifter. The first one suppresses the phase variations introduced by the optical fiber required for the practical implementation of our phase shifter, and the second one also diminishes the fluctuations due to thermal variations of the AOMs. While the second setup is more reliable, this comes at the expense of having additional frequencies present at the output. We showed with our setup that a phase shift of $2\pi$ can be achieved with a \SI{10}{\meter} fiber by varying the AOM frequency $\fAOM=\fS/2$ in a range smaller than $\SI{10}{\mega\hertz}$. A greater phase shift of several cycles can be achieved by using a longer optical fiber or a greater frequency shift.

The presented technique opens a new pathway in phase control methods. By leveraging the frequency of an electronic signal to determine the phase shift, this technique harnesses the unparalleled precision inherent in frequency synthesizers.



\begin{backmatter}
\bmsection{Acknowledgment}
The authors would like to thank Luis A. Orozco for helpful discussions. 
\bmsection{Funding}
This work was supported by project UNAM-PAPIIT IN115523, UNAM-CIC LANMAC program and CONACyT Ciencia Básica grant A1-S-29630. 
E. E-R., L. U-R., E. G. A-T. and A. H-L. acknowledge fellowships by CONAHCyT.
\bmsection{Disclosures}
The authors declare no conflicts of interest.
\bmsection{Data availability} Data underlying the results presented in this paper are not publicly available at this time but may be obtained from the authors upon reasonable request.
\end{backmatter}
\\


\bibliography{references}


\end{document}